\documentclass[final,a4paper,12pt]{elsarticle}
\usepackage{hyperref}
\usepackage{graphicx}

\begin{document}

\begin{frontmatter}
\title{Condensation of laser produced gold plasma during expansion and cooling in water environment}

\author[1,2]{Yu.~V.~Petrov}
\author[1,3]{N.~A.~Inogamov*}\ead{nailinogamov@gmail.com}
\author[3,1]{V.~V.~Zhakhovsky}
\author[1]{V.~A.~Khokhlov}

%\authormark{AUTHOR ONE \textsc{et al}}

\address[1]{Landau Institute for Theoretical Physics of RAS, 1-A Akademika Semenova
av., Chernogolovka, Moscow Region, 142432, Russia}
\address[2]{Moscow Institute of Physics and Technology, 9 Institutskii per., Dolgoprudny,
Moscow Region, 141700, Russia}
\address[3]{Dukhov Research Institute of Automatics, 22 Sushchevskaya st., Moscow, 127055, Russia}

%\presentaddress{This is sample for present address text this is sample for present address text}

\begin{abstract}
The ecologically best way to produce nanoparticles (NP) is based on laser ablation in
liquid (LAL). In the considered here case the LAL means that a gold target is irradiated through transparent
water. During and after irradiation the heated material from surface of a target forms a plume which expands
into liquid.
 In this paper we study a reach set of physical processes mixed with complicated hydrodynamic phenomena
   which all accompany LAL.
 These theoretical and simulation investigations are very important for practical applications.
 Laser pulses with different durations $\tau_L$ covering 5-th orders of magnitudes range from 0.1 ps to 0.5 ns
   and large absorbed fluences $F_{abs}$ near optical breakdown of liquid
     are compared.
 It is shown that the trajectory of the contact boundary with liquid at the middle and late stages
  after passing of the instant of maximum intensity of the longest pulse
   are rather similar for very different pulse durations (of course at comparable energies $F_{abs});$
     we consider the pulses with a Gaussian temporal shape $I\propto \exp(-t^2/\tau_L^2).$
 We follow how hot (few eV range) dense gold plasma expands, cools down, intersects a saturation curve,
   and condenses into NPs.
 These NPs appear first inside the water-gold diffusively mixed intermediate layer
   where gold vapor has the lowest temperature.
 Later in time pressure around the gold-water contact drops down below critical pressure for water.
 Thus NPs find themselves in gaseous water bubble
   where density of water gradually decreases to $10^{-4}-10^{-5}$ g/cm$\!^3$
     at the instant of maximum expansion of a bubble.
\end{abstract}

%\keywords{keyword1, keyword2, keyword3, keyword4}

\end{frontmatter}

\section{Introduction}

 Technological aspects and importance of wide variety of nanoparticles (NP) produced by laser ablation in liquids (LAL)
   were recently described in large review papers \cite{Stephan:2017.review,XIAO:2017}.

 As usual, engineering challenge coming from real life brings a lot of physical problems belonging to separate directions.
 The present work on NP/LAL is significant in the fundamental sense because it links together these scientific directions.
 These directions include propagation of shocks in condensed media,
   first order equilibrium and non-equilibrium phase transitions (melting/solidification; evaporation/condensation)
    in hydrodynamically evolving gold and water,
      atomic solvability of gold in water is important,
       it is necessary to know transport coefficients,
         e.g., diffusion coefficients defining intermixing between contacting components.
 We consider transformation of dense hot plasma states of gold
   into rather low temperature two-phase vapor-liquid mixture with growing NPs.
 The transformation takes place in liquid surrounding.
 Both gold and water transit from overcritical densities to subcritical densities
   during heating, expansion, and cooling processes.
 The pass through near-critical densities causes strong increase in compressibility.
 Gold first transits to the soft states, later in time water also follows this way.
There are three ranges relative to the degree how rigid/soft matter is.
In condensed states the index $\gamma=d(\log(p))/d(\log(\rho))$ is high (e.g, $\sim 3$ and more);
 in gaseous states above a critical point a thermal contribution into pressure becomes significant and $\gamma$ is $\sim 1.4-2;$
 in two-phase liquid-vapor mixture we have the lowest values for $\gamma\approx 1+\epsilon,$ $\epsilon\ll 1.$

\begin{figure}       %  ---------------------------------------------  РИС. 1
   \centering   \includegraphics[width=0.9\columnwidth]{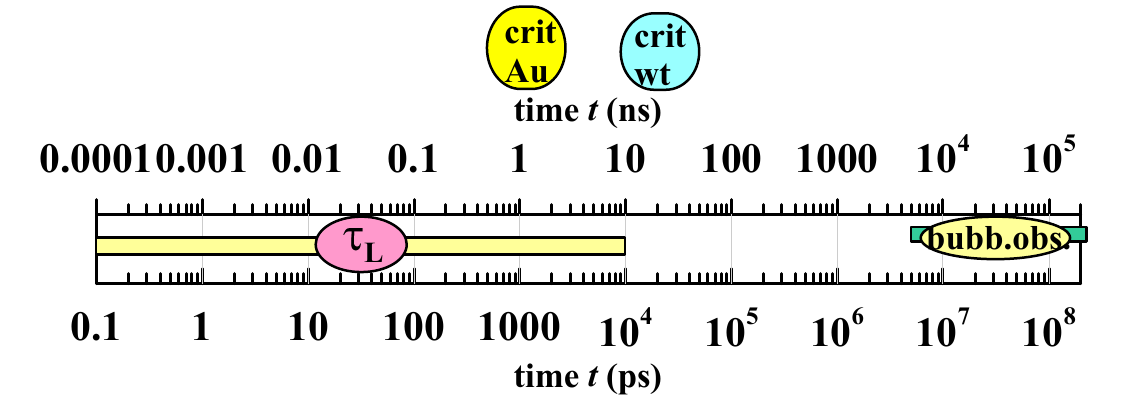}
\caption{\label{ris:01-hierarchy}
The hierarchy of scales for the chain of processes triggered by laser irradiation of a gold target through water.
Pulse durations $\tau_L$ usually used in LAL are from $\sim 0.1$ ps to $\sim 10$ ns.
High speed camera records appearance, expansion, and collapse of bubble over the range marked as "bubb. obs.".
We refer here for recent data on bubble observations published in \cite{Amans:2016:APL}.
We see that observations are difficult in the long era preceding to bubble formation.
Pressure $p_{CB}(t)$ near the Au-water contact boundary (CB) decreases with time.
First it drops below the critical pressure for gold and next below the critical pressure for water. Corresponding stages are marked as "crit Au" and "crit wt".
    }  \end{figure}

 Formation of a bubble is consequence of softening of water.
 Evolution range of typical LAL is shown in Fig. \ref{ris:01-hierarchy}.
 It covers very wide range of time scales.
 In the case of femtosecond action this range consists from 9-th orders of magnitudes
   starting from duration of a pulse $\tau_L$
    and finishing at the instant when a bubble achieves its first expansion maximum.
 And only last two orders are carefully observed by ultrafast camera, see recent detailed studies in \cite{Amans:2016:APL}.
 It is convenient to describe initial stages in picoseconds in the cases of femtosecond and picosecond actions.
 Middle stage is measured in nanoseconds, while the final stages,
  when a semi-spherical bubble is formed and begin to expands, run along the microsecond time scales.
 These ps-ns-$\mu$s ranges are illustrated in Fig. \ref{ris:01-hierarchy}.

% 1-4

\begin{figure}       %  ---------------------------------------------  РИС. 2
   \centering   \includegraphics[width=0.7\columnwidth]{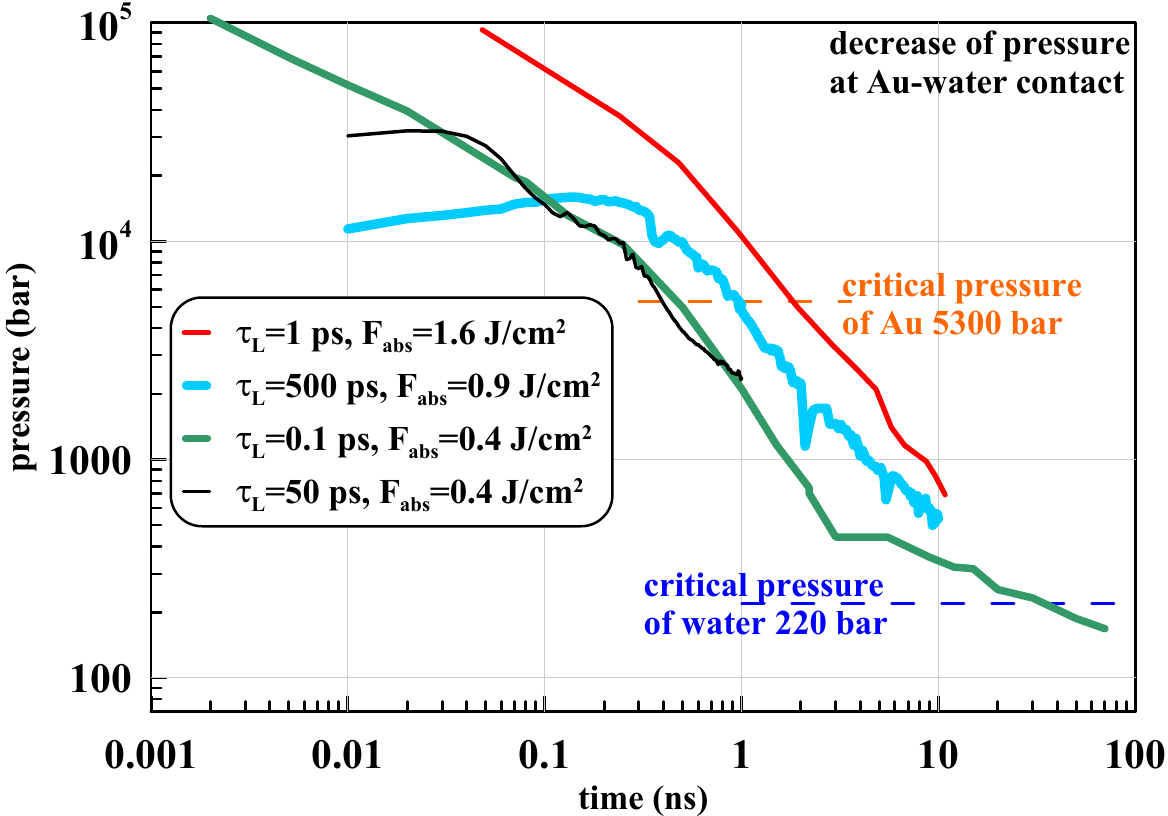}
\caption{\label{ris:02-pCB-t} Pressure at a contact.
The upper curve (the red curve) is taken from molecular dynamics (MD) simulation
  which will be described in the present paper below.
Other three dependencies are results of hydrodynamic simulations.
Unloading of shock compressed water as shock in water runs away decreases contact pressure $p_{CB}(t).$
Thus it drops first below critical pressure for gold and after that below critical pressure of water.
These stages are marked in Fig. \ref{ris:01-hierarchy} above as "crit Au" and "crit wt".
All these simulations were done for the gold-water case.
 The evolution corresponding to the shot with $F_{abs}=0.4$ J/cm$\!^2$ and $\tau_L=0.1$ ps in Fig. \ref{ris:02-pCB-t}
   is taken from the paper \cite{INA.jetp:2018.LAL}.
    }  \end{figure}

 Pressure at a contact boundary $p_{CB}(t)$ is an important function of a problem.
 After finishing of a pulse it gradually decreases with time as it is shown in Fig. \ref{ris:02-pCB-t}.
 During heating by a laser pulse the pressure $p_{CB}$ increases achieving its maximum value at the end of a pulse.
 This is definitely seen for the pulses with durations $\tau_L$ equal to 50 ps and 500 ps.
 It is important that the descending part of the function $p_{CB}$ below $\sim 1$ GPa is universal
   relative to duration of a pulse.
 This follows from simulations, see examples presented in Fig. \ref{ris:02-pCB-t}.
 The function $p_{CB}(t)$ is approximately described as a power law decrease in time
   with a coefficient proportional to energy absorbed during a laser action:
\begin{equation}\label{eq:01-pCB}
P_{CB}(t) = 5\cdot 10^3 F_{Jcm2}/t_{ns}^{1.05}\,{\rm [bar]}=0.5 F_{Jcm2}/t_{ns}^{1.05}\,[\rm{GPa}],
\end{equation}
 where $F_{Jcm2}=F_{abs}/1$[J/cm$\!^2$] is absorbed energy, $t_{ns}=t/1$[ns] is time in nanoseconds
   reckoned from a maximum of a Gaussian pulse $I=I_0\exp(-t^2/\tau_L^2).$

% 1-5

 There are five expansion stages.
 The first stage corresponds to heating of a target by a laser pulse.
 It is well seen in Fig. \ref{ris:02-pCB-t} for shots with pulse durations $\tau_L$ equal to 50 ps and 500 ps.
 For ultrashort pulses the pressures at the first stage are high, thus they are not shown in Fig. \ref{ris:02-pCB-t};
   for the shot with $F_{abs}=0.4$ J/cm$\!^2$ and $\tau_L=0.1$ ps (the green curve in Fig. \ref{ris:02-pCB-t})
     the first stage is presented in details in \cite{INA.jetp:2018.LAL}.
 The second stage relates to transition from the first stage to the third stage
   where $p_{CB}(t)$ behave as $\propto 1/t,$ see (\ref{eq:01-pCB}).

\begin{figure}       %  ---------------------------------------------  РИС. 3
   \centering   \includegraphics[width=0.75\columnwidth]{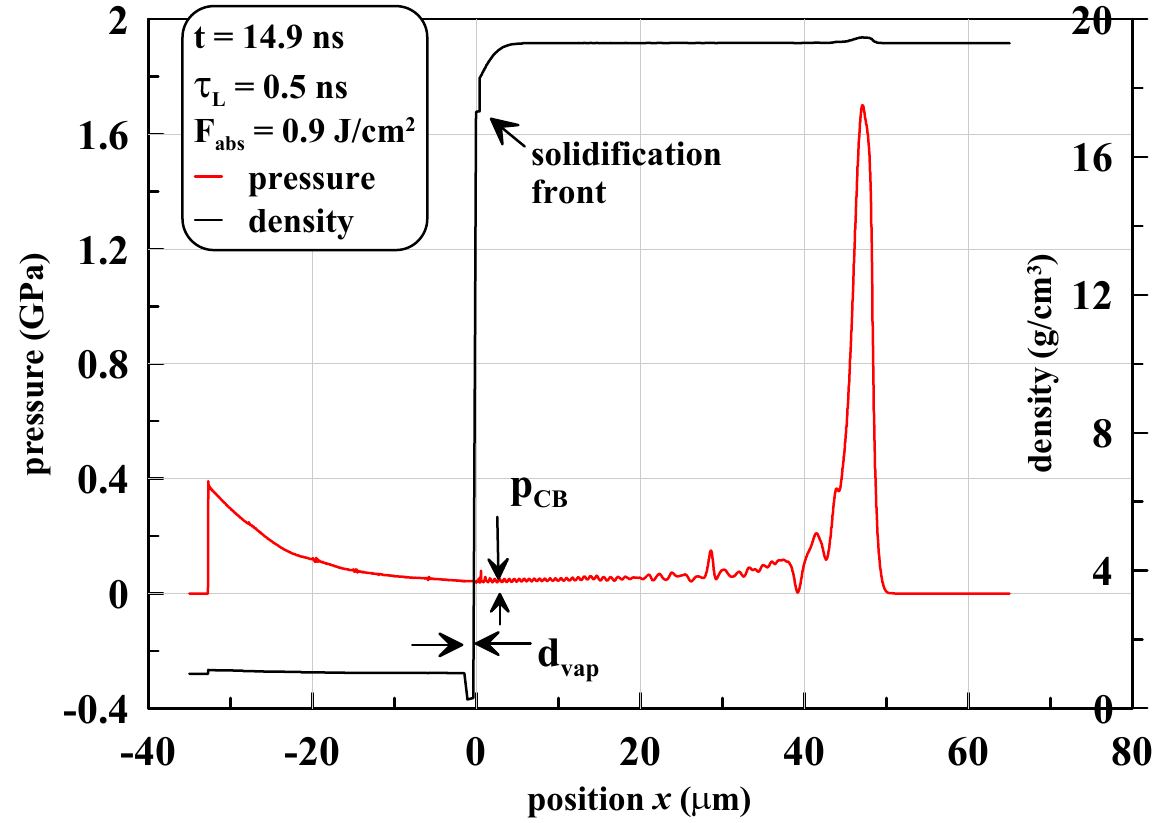}
\caption{\label{ris:03-p(x)} Pressure and density profiles showing two shocks and near contact layer between them.
Pressure $p_{CB}$ is contact pressure plotted in Fig. \ref{ris:02-pCB-t} as the blue curve.
The arrows $d_{vap}$ mark the instant boundaries of a layer of a gaseous gold (vapor).
The tip of the left arrow $d_{vap}$ denotes the Au-water contact boundary (CB).
The tip of the right arrow $d_{vap}$ marks a boundary between gaseous and liquid gold.
    }  \end{figure}

\begin{figure}       %  ---------------------------------------------  РИС. 4
   \centering   \includegraphics[width=0.75\columnwidth]{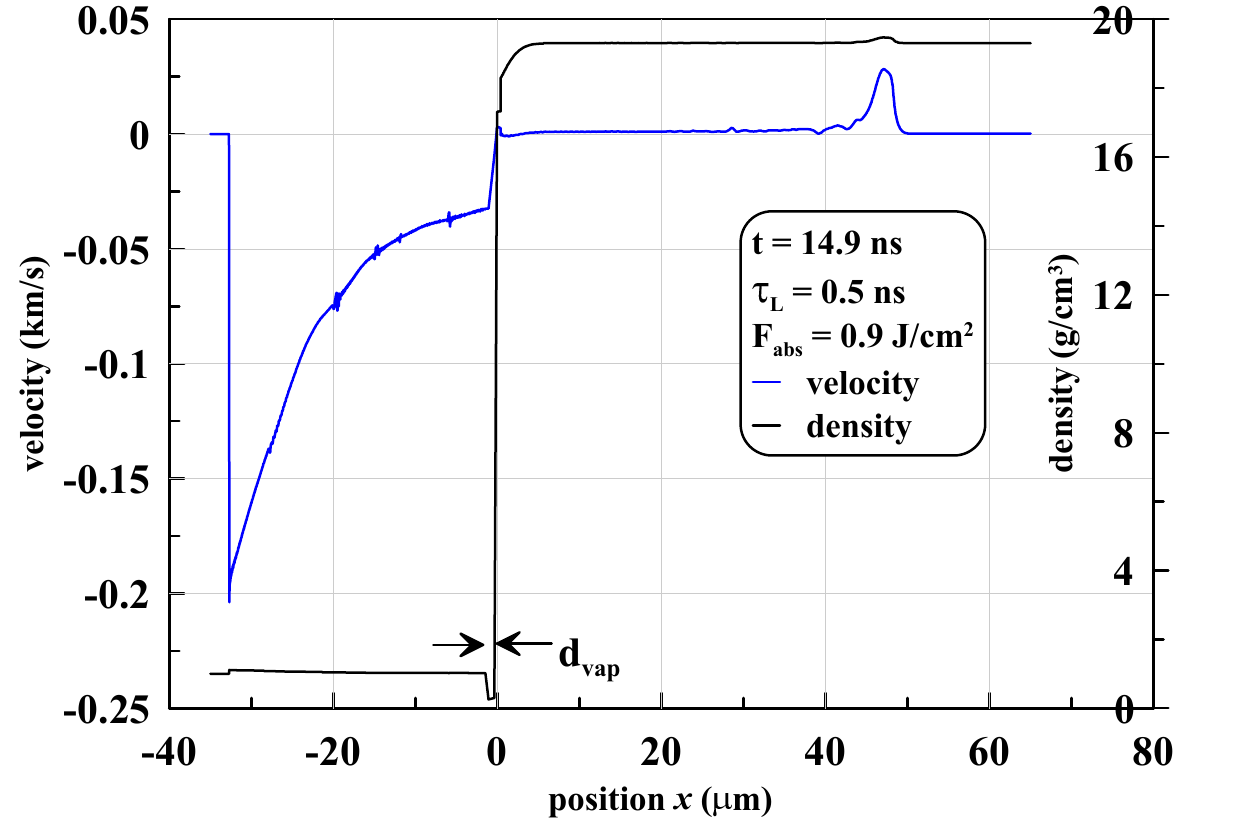}
\caption{\label{ris:04-u(x)} Instant profiles of velocity and density.
We see that condensed gold (liquid and solid, see the arrow "solidification front" in previous Fig. \ref{ris:03-p(x)})
is motionless except the layer occupied by a shock wave moving through bulk gold to the right side.
Thus motionless is the right boundary of the vaporized gold layer $d_{vap}.$
    }  \end{figure}

 % 1-6

 The second stage almost disappears in the case of nanosecond pulse shown by the blue curve in Fig. \ref{ris:02-pCB-t}.
 But it is well presented for shorter pulses, even for the pulse with $\tau_L=50$ ps.
 At the second stage the pressure $p_{CB}(t)$ decays more slow
   than the dependence $\propto 1/t$ (\ref{eq:01-pCB}) valid at the third stage.
 This is because there are nucleation, rupture, and foam formation in the cases with ultrashort action
   and also in the case with multi-picosecond pulse $\tau_L=50$ ps.
 Internal layers of foam move faster than the "atmosphere" created by the deceleration of the contact by water.
 Resistance of the water to expansion of a gold plume is described in the paper \cite{INA.jetp:2018.LAL}
   devoted to explanations of this phenomenon.
 The inflow of foam into a contact plug or "atmosphere" was observed also in papers
       \cite{POVARNITSYN:2013,SHIH20173,LZ+Stephan:2018.LAL}
          where ultrashort pulses and moderate energies $F_{abs}$ were considered.
 Thus ram pressure of foam "accreting" (see \cite{INA.jetp:2018.LAL}) onto atmosphere contradicts to deceleration of a contact
   by water resistance.
 Appearance of the ram pressure is a result of the inflow of momentum into atmosphere.
 This inflow partially compensates deceleration and partially compensates decay of contact pressure.

 % 1-7

 Therefore thanks to this compensation
   the contact pressure decreases more slowly at the second stage relative to the third stage.
 At the third stage the ram support from the foam side finishes
   and degree of deceleration of a contact by liquid increases.
 Or this support doesn't exist as in the case of nanosecond pulse shown by the blue curve in Fig. \ref{ris:02-pCB-t}.
 We don't see nucleation and formation of foam in the nanosecond action, because heating is slow,
   much more slower relative to the acoustic time scale $t_s=d_T/c_s\sim 30$ ps,
    where $d_T$ is thickness of heat affected zone or ablated layer, $c_s\approx 3$ km/s is speed of sound in gold.

 Very thick foamy layer is formed by an ultrashort pulse at large absorbed energies $F_{abs}.$
 Foam inflates greatly (in the case of expansion into vacuum) and becomes many times thicker
   than thickness of a heat affected zone or thickness of ablated layer (thus volume fraction of liquid in foam becomes small).
 Inflation process lasts longer and longer time as fluence $F_{abs}$ increases \cite{Vorobyev:2017};
   up to hundreds of nanoseconds, see \cite{Vorobyev:2017}.
 In the case with transparent liquid the liquid strongly resists to expansion of foam.

\begin{figure}       %  ---------------------------------------------  РИС. 5
   \centering   \includegraphics[width=0.75\columnwidth]{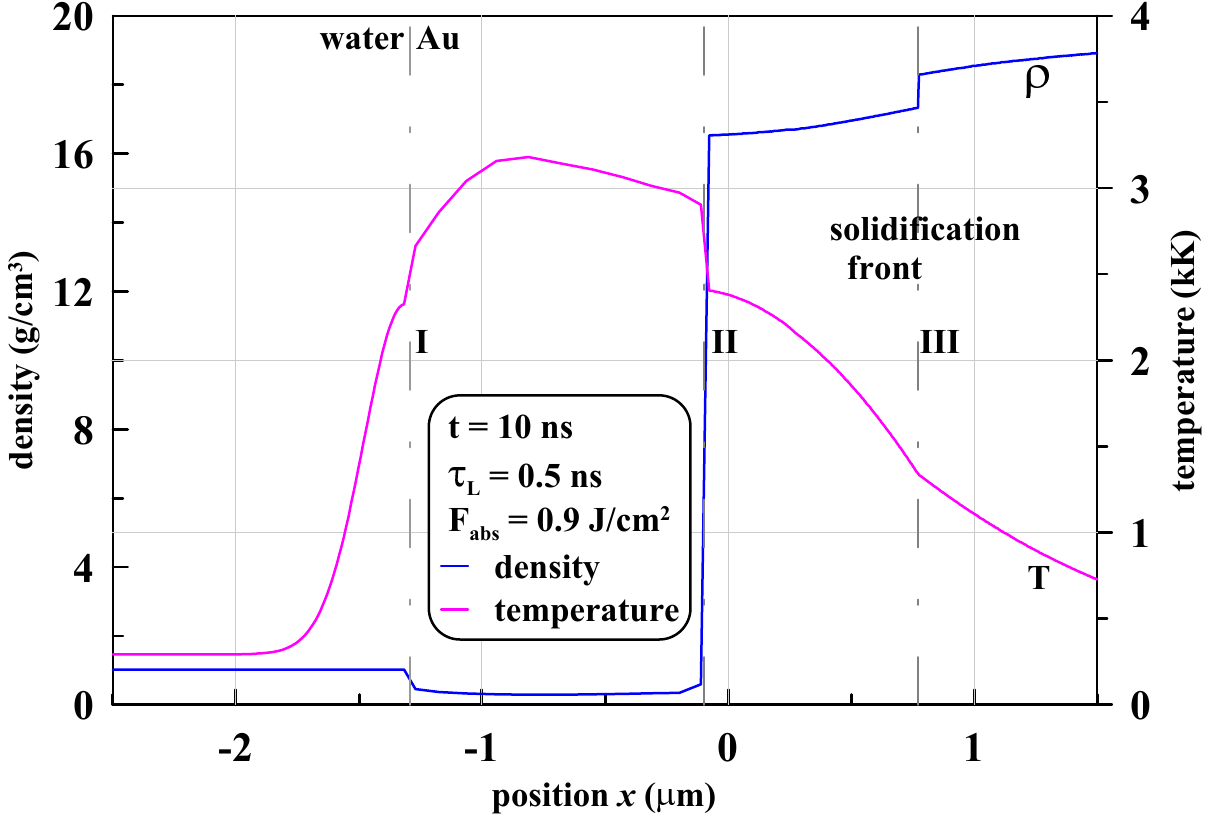}
\caption{\label{ris:05-contactLayer} Structure of a contact layer
 according to approximations used in our hydrocode;
   shortcomings and advantages of these approximations are discussed in the text.
The layer consists from gaseous gold (located between I and II)
 and a zone of hot water near the contact boundary marked as I.
Our hydrocode is based on splitting of mass along the spatial axis into a Lagrangian mesh.
The code doesn't include mutual diffusion of gold atoms and water molecules.
Therefore the contact boundary I remains sharp.
The number III marks solidification front.
See text where freezing process is described.
    }  \end{figure}

 % 1-8

 At the third (III) stage (\ref{eq:01-pCB}) the contact pressure is defined by pressure of approximately adiabatic gaseous gold
  in one-phase states above the condensation curve
    and by space left to gaseous gold by retreating water
      (the drop is a result of expansion of a layer filled by gaseous Au).
 The change to the fourth stage of evolution begins when gold transits from one-phase states to two-phase states.
 Gold intersects the condensation (or saturation, or equilibrium) curve
   and begins to form the condensation clusters (few atoms together) growing into nanoparticles (hundreds and more atoms together).
 At the fourth (IV) stage the compressibility of gold achieves its maximum.
 It is larger than at the third stage where gold was in one-phase states;
  let's mention that compressibility of one-phase gaseous gold is larger than compressibility of dense gold
     with densities higher that critical density $\rho_{crit};$ for gold $\rho_{crit}\approx 5.3$ g/cm$\!^3.$

\begin{figure}       %  ---------------------------------------------  РИС. 6
   \centering   \includegraphics[width=0.75\columnwidth]{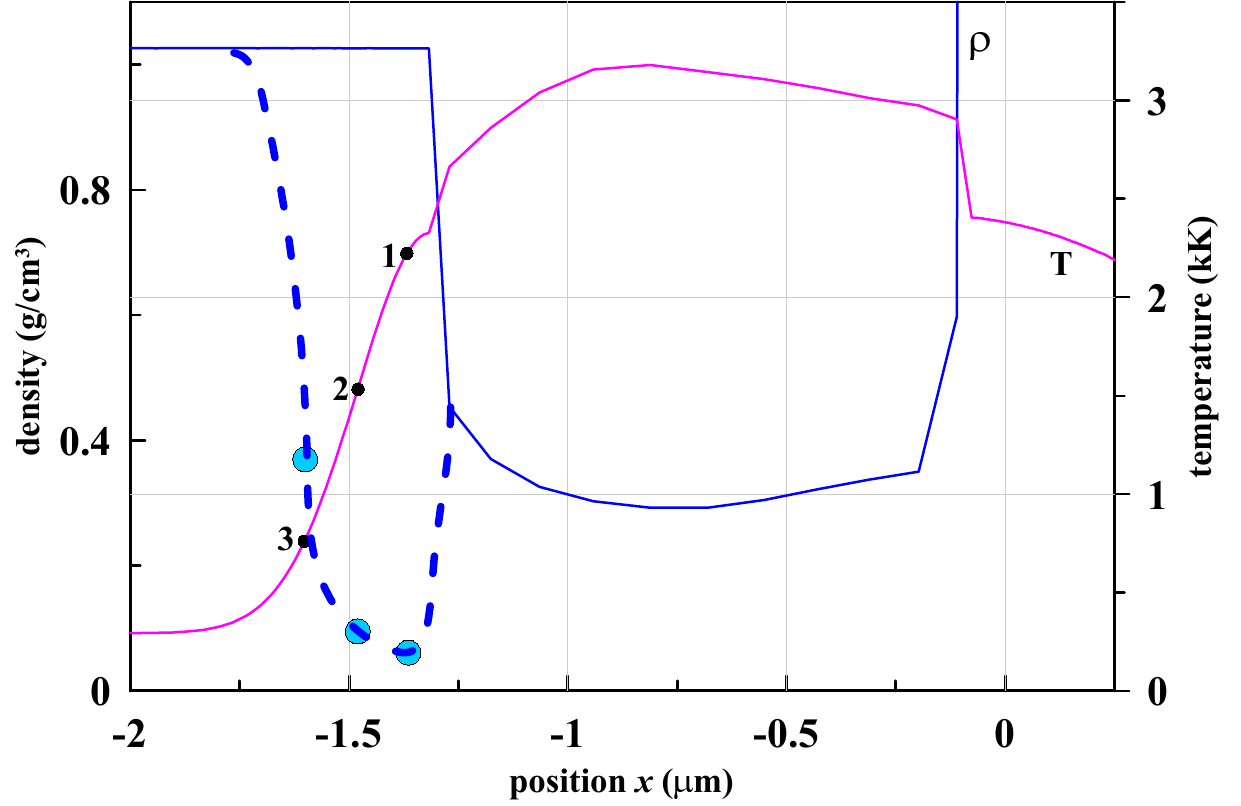}
\caption{\label{ris:06-contactLayer} This is an enlarged view of the contact layers
   taken from previous Fig. \ref{ris:05-contactLayer}.
 The structure consists from a layer of gaseous gold and a layer of hot water adjacent to gaseous gold.
 Correction to the structure of the contact layers is introduced.
 It comes from volume thermal expansion of water under given pressure thanks to rise of temperature.
 Uncorrected density profile is shown by the continuous blue curve.
 The corrected density profile has smaller density inside the layer of hot water.
 Decrease of density is shown by the interval of the blue dashed curve passing through three blue filled circles.
    }  \end{figure}

 % 1-9     pro 3-yu kak predverie 4-i

 At the fourth stage the dependence $p(\rho, s)$ for gold becomes softer
  (an adiabatic index drops to $\gamma\approx 1)$
    therefore decrease of contact pressure with time as gold expands becomes more slow relative to the rather fast decay
      $p_{CB}\propto 1/t,$ see (\ref{eq:01-pCB}) and Fig. \ref{ris:02-pCB-t}; here $s$ is entropy.
 The kink in the $p_{CB}(t)$ dependence due to transit through a condensation curve
   is definitely seen at the green curve in Fig. \ref{ris:02-pCB-t}.
 Obviously, the kink marking the change of stages III and IV should be below critical pressure for gold.
 The position of the kink depends on temperature (it is better to say on entropy $s)$ of gaseous gold -
   the higher $s,$ the larger the expansion and longer time is necessary for the III$\to$IV transit.
 At the fourth stage the pressure of high entropy gold
    equals to saturation vapor pressure $p_{sat}(T)$ at given temperature.
 This means that intensive formation of nanoparticles begins thanks to condensation process.
 The fifth stage begins when contact pressure will be defined by pressure of gaseous water, see next Sections.

 % 1-10  обе ув вместе  сказать, что донное ==нижн давл не нуль! Это из-за противодейств среды

 Figures \ref{ris:03-p(x)} and \ref{ris:04-u(x)} show propagation of shocks in water and in gold.
 We see how far are the shocks relative to very thin (at the spatial scale connected with shocks) layers
   of vaporized and molten gold $d_{vap}\sim 1$ $\mu$m.
 Still the one-dimensional approach is valid because a path passed by a water shock
   is much less than radius $R_L$ of a laser beam.
 We will orient on the value $R_L\approx 250$ $\mu$m given in paper \cite{Amans:2016:APL}.
 In the paper \cite{Amans:2016:APL} it was defined by a crater diameter.
 The shock in gold is insignificant for dynamics of vaporized gold and water.

 % 1-11

 Finite pressure $p_{CB}$ in the intermediate region between shocks (see Fig. \ref{ris:03-p(x)})
   is formed thanks to water resistance to ablative expansion of hot matter.
 In the case with expansion in vacuum the pressure at the bottom of the future crater is zero at the rather late stage
   shown in Figures \ref{ris:03-p(x)} and \ref{ris:04-u(x)}.
 Let us mark also that at the stage shown in these Figures mainly the vaporized gold is expanding,
   compare gradients of velocity near the boundary of condensed gold.
 The last sentence is significant.
 It will be discussed and made more exact below.
 Large volume expansion of thin, high entropy gold and water layers (as pressure decreases) will be opposed to
   small volume expansion of surrounding condensed gold and cold water.

 % 1-12 оглавление следщх разделов
are
 % In the next Chapters

 \section{Structure of high entropy layer according to hydrodynamic simulations}

 % 2-1  T//kappa, kot ne trogaet Hug

 In Figures \ref{ris:03-p(x)} and \ref{ris:04-u(x)} an integral picture at a rather late instant of time is shown.
 It includes the whole Au-water layer between two shocks.
 The thin layer around $d_{vap}$ in Figures \ref{ris:03-p(x)} and \ref{ris:04-u(x)} is the most interesting place
   for technologies of nanoparticles production.

 % 2-2

 Fig. \ref{ris:05-contactLayer} presents structure of a contact layer $d_{vap}$ obtained by using of a hydrodynamic code.
 This code was described in \cite{INA.jetp:2018.LAL}.
 Modification introduced into the code in the present paper relates to thermal conductivity of water.
 Thermal conduction in water was neglected in \cite{INA.jetp:2018.LAL}.
 A layer of hot water appears thanks to conduction, see Fig. \ref{ris:05-contactLayer}.
 Of course, water conduction is weak in comparison with condensed gold, but it is important because we want to understand how a bubble is formed
   and how gold nanoparticles and atomic gold are mixed with gaseous water inside a bubble.

 % 2-3

 Indeed, there are two sources of heating of water.
 The first one is connected with dissipation in shock (entropy trace after shock).
 As we will see below this source is weaker than the conductive source (the second source of heating);
   dissipation power nonlinearly depends on Mach number of shock; for long pulses the dissipation is negligibly small.
 Thus conduction is important.
 But the first source (in the cases with short pulses) creates a wider layer of heated water than the second source, see molecular dynamics simulations below.

 % 2-4  kappa H2O, kotor v raschete = zavishen. ne stal eto obsuzhdat'

 In the particular simulation presented in Figures \ref{ris:03-p(x)}-\ref{ris:05-contactLayer} the coefficient $\kappa$
  of thermal conduction of water was taken equal $\kappa_{eff}=6$ W/m/K; $\kappa=318$ W/m/K for gold in normal conditions;
    $\kappa$ for gold in wide range of temperatures and densities is taken from
      \cite{Petrov:2013,Petrov:Au:kappa:2015,Migdal:2T.eos.kappa:2017}.
 Thickness of a heated water layer is $d_{wt}(t=10\,{\rm [ns]}) = 0.22$ $\mu$m (FWHM)
   at the instant $t=10$ ns shown in Fig. \ref{ris:05-contactLayer}.
 History of heating of water from gold and propagation of heat in water is complicated
   because thermal conductivities $\kappa$ of gold and water significantly change
     thanks to wide variations in temperature and density.
 Coefficient $\kappa$ of gold in gaseous states in the layer between boundaries I and II in Fig. \ref{ris:05-contactLayer}
   is very small.
 Thus heat absorbed by water is accumulated during rather early stages (before the instants shown in Figures \ref{ris:03-p(x)}-\ref{ris:05-contactLayer})
    when gold near a contact was dense and better conducting.

 % Pressure at the instant shown in Fig. \ref{ris:05-contactLayer} and Fig. \ref{ris:06-contactLayer} is 600 bar.
 % This pressure is much higher than saturated vapor pressure $p_{sat}^{Au}(T)$ for gold at temperatures $\approx 3$ kK:
 %   $p_{sat}^{Au}(T=3\, [{\rm kK}])$ is less than 1 bar, see Fig. (4) in \cite{INA.jetp:2018.LAL}.
 % Spatial expansion of the layer I-II is very slow (this expansion defines adiabatic cooling),
 %   conduction loses are suppressed,
 %     thus for long time gold remains under pressure overcoming its saturation pressure, and condensation is delayed.

\begin{figure}       %  ---------------------------------------------  РИС. 7
   \centering   \includegraphics[width=0.75\columnwidth]{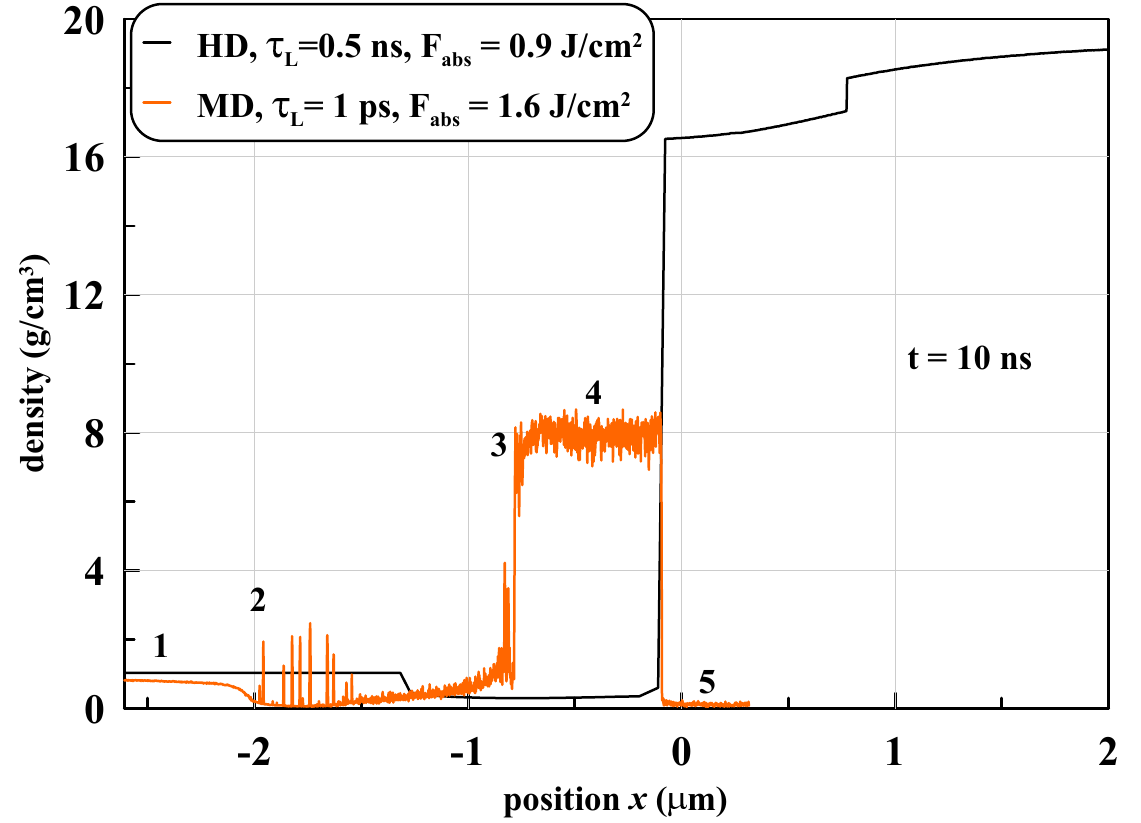}
\caption{\label{ris:07-HD-MD-densities} Water heated by a shock passed through water some time ago is denoted as "1".
"2" is the left edge of the diffusion layer located between 2 and 3. The needles near the 2 and 3 are the gold clusters.
Thus condensation develops heterogeneously starting from the edges. Formation of the clot 4 and the rarefied region 5
 is discussed in text.
    }  \end{figure}
 % 2-4.1

 In this paper to describe heating of water we use effective coefficient $\kappa_{eff}=6$ W/m/K.
 Coefficient $\kappa_{eff}$ is $0.3-0.75$ W/m/K in {\it liquid} water in a pressure range 1-1000 bars, see \cite{engineeringtoolbox};
   transport in liquid and gaseous water is very different.
 Thermal diffusivity $\chi=\kappa/C$ is used when solving heat conduction equation $T_t=\chi T_{xx},$
   here $C$ is heat capacity per unit of volume.
 Thermal diffusivity $\chi$ is $\sim 0.001$ cm$\!^2/$s for liquid water.
 Coefficient $\chi$ for gaseous overcritical water is roughly $\chi\sim 0.03$ cm$\!^2/$s
   for $p=100$ bar, $T=1.2$ kK \cite{engineeringtoolbox}.
 This coefficient is 30 times larger than the coefficient for liquid water.
 But heat capacity per molecule is decreased from $\approx 9 k_B$ in liquid to $\approx 3 k_B$ in gas.
 Therefore we use as an approximate effective thermal conductivity for water the value $\kappa_{eff}=6$ W/m/K.
 Additional information following from molecular dynamics simulations is given below.

 % 2-4.2

 Redistribution of heat thanks to conduction before the geometrical transition of a plane (disk type) layer of hot water
   in semi-spherical bubble shape continues up to few microseconds \cite{Amans:2016:APL}.
 % ing of water from gold continues up to the stage when high entropy water transits into the bubble formation stage.
 % Estimates of the time necessary for this transition give $t_{gasWt}\sim 10^2-10^3$ ns.
 % After transition the conductive accumulation of entropy in the water layer gradually ceases.
 %Thus final thickness of a layer of hot water before it transits to a semi-spherical shape is
 Thickness of a layer of high entropy water before transition to a semi-spherical shape is
 \begin{equation} \label{eq:02-dWaterFinal}
   d \sim d_{wt}(t=10\,{\rm [ns]}) \sqrt{t_{geom}/(t=10\,{\rm [ns]})} \sim 2-5 \, \rm{[\mu m]},
 \end{equation}
   where $t_{geom}$ is 1-5 $\mu$s.
 The layer (\ref{eq:02-dWaterFinal}) is very thin in comparison with typical diameters of a laser beam $\sim 500$ $\mu$m.
 Hydrodynamic expansion velocities overcome velocities of heat conduction spread $v_T\sim\sqrt{\chi/t}$
  $(v_T\sim 1$ m/s for $\chi=0.03$ cm$\!^2/$s and $t=1$ $\mu$s) during and after transition.

 %  sloi progreva tonok
 % progrev do perekhoda v gas-vodu = slom=kogda? posm ris 3

               %%   chto kappa Au soglasno nashim pred rabotam=kakim? vot oni:
 % d:\c\D\ORG\PUBL\OTTISKI\2018\Elbrus17\Petrov_2018_J._Phys.__Conf._Ser._946_012096.pdf  ищу каппа Ау. здесь Ал
 % d:\c\D\ORG\PUBL\OTTISKI\2017\1702.00825.pdf  есть каппа Ау,но при норм плотности   *
 % d:\c\D\ORG\PUBL\OTTISKI\2016\Ap.Ph.A.2016_2Tkappa+alpha=Cu_RSCF.pdf           каппа медь
 % d:\c\D\ORG\PUBL\OTTISKI\2016\Elbrus\JPCS_774_1_012103-Petrov,Migdal-2016.pdf  каппа медь. есть флы ЮП для завсти от плотн
 % d:\c\D\ORG\PUBL\OTTISKI\2016\2T.kappa.Cu\2T_kappa,sigma_Cu_Petrov_jetpLett2016.pdf
 % d:\c\D\ORG\PUBL\OTTISKI\2015\Elbrus\Migdal_2T_Elbrus-2015_bez.blagdrnst.pdf
 % d:\c\D\ORG\PUBL\OTTISKI\2015\Elbrus\Petrov_2T_Elbrus_blagdrnst=1078.pdf   *

 % еще ссылки на Ау-вода =
 % d:\c\D\ORG\PUBL\OTTISKI\2018\1803.07343.pdf
 % d:\c\D\ORG\PUBL\OTTISKI\2018\AIP-SCCM-2017\Las.Abl.Liq_AIP(2018).pdf

 % 2-5   gde Hug ??  Hug--> rhoWater = 1 near CB

 Hydrocode used employs a simplified version of equation of state for water.
 It is taken from a Hugoniot (or shock) adiabatic curve,
   see \cite{INA.jetp:2018.LAL,INA.AIP:2018.LAL,INA.arxiv:2018.LAL}.
 This is an one-argument function giving dependence of pressure in water as a function of water density: $p_{Hug}(\rho),$
   see (12) in \cite{INA.jetp:2018.LAL}.
 Temperature field calculated in water by the hydrocode doesn't influence the dependence $p_{Hug}(\rho)$
   and thus the density profiles $\rho(x,t)$ in water.
 Therefore density of water is approximately 1 g/cm$\!^3$ near the contact I in Fig. \ref{ris:05-contactLayer}
  and in Fig. \ref{ris:06-contactLayer} (the blue continuous curve);
   the adiabatic curve $p_{Hug}(\rho)$ (12) in \cite{INA.jetp:2018.LAL} returns to uncompressed water in normal conditions
     when pressure drops down significantly below bulk modulus $1.5$ GPa for water.
 This is the reason why the blue continuous curve presenting density profile
   in Figures \ref{ris:05-contactLayer} and \ref{ris:06-contactLayer}
     has density $\approx 1$ g/cm$\!^3$ in water close to the contact I.

 % 2-6    pereschet :  p-shiroko, T==priblizhenno=ne zavisit
 %       каппа в 0.5нс.ее отключка от ро. р и скор ок. тонк слой гор воды не описывается. р(ро)=Гюгонио

 Correction to density of water due to it heating is given in Fig. \ref{ris:06-contactLayer} by the dashed blue interval.
 Let's say how we have calculated this correction.
 First, take pressure $p=600$ bar in the contact region for the instant $t=10$ ns shown in Fig. \ref{ris:06-contactLayer};
   see Figures \ref{ris:02-pCB-t} and \ref{ris:03-p(x)} where pressures are present.
 We suppose that this pressure moderately depends on the correction.
 Second, choose three points 1, 2 and 3 at the temperature profile of hot water in Fig. \ref{ris:06-contactLayer}.
 Take temperatures $T_j,$ $j=1,2,3$ in these points.
 Third, use equation of state for water $p(\rho,T)$ to calculate densities $\rho_j$ for these pairs $(T_j, p)$
   from equations $p(\rho_j, T_j)=600$ bar.
 We use equation of state for water given in paper \cite{NB:2011}.

 %   ссылка NB:2011  из биб файла   Au-glass_50ps_1.8.18.bib

 % 2-7

 We plot values of obtained densities $\rho_j$ at the same positions $x_j$
   where the points 1, 2 and 3 in Fig. \ref{ris:06-contactLayer} are placed.
 Thus the blue dashed interval going through the three blue circles $x_j,\rho_j$ in Fig. \ref{ris:06-contactLayer} appears.
 We see that water in the hot layer also (as gold) very significantly expands even at rather significant load - 600 bar.
 At the instant $t=10$ ns the pressure in water 600 bar is larger than critical pressure 220 bar for water.
 In the dashed interval the water is in gaseous overcritical state.
 Below we will return to consideration of decrease of pressure to low values much smaller than 220 bar.
 This will be prolongation of the dependencies in Fig. \ref{ris:02-pCB-t} down to 0.1-1 bar.
 %% будет ли сие? будет ли сферизац и ур РП?

 % 2-8    See text where freezing process is described. melt t-solid=?*

 % \section{Freezing of condensed gold and formation of surface structure}

 Thermal conductivity in condensed gold remains high.
 Thus the layer of molten gold seen in Figures \ref{ris:03-p(x)}-\ref{ris:05-contactLayer}
   will be solidified in next few tens of ns: $\sim d_{molten}^2/\chi_{Au}.$
 Surface structures are formed during solidification.
 Vapor plume in the case of long (ns) pulses weakly affects these structures during solidification.

\begin{figure}       %  ---------------------------------------------  РИС. 8
   \centering   \includegraphics[width=0.75\columnwidth]{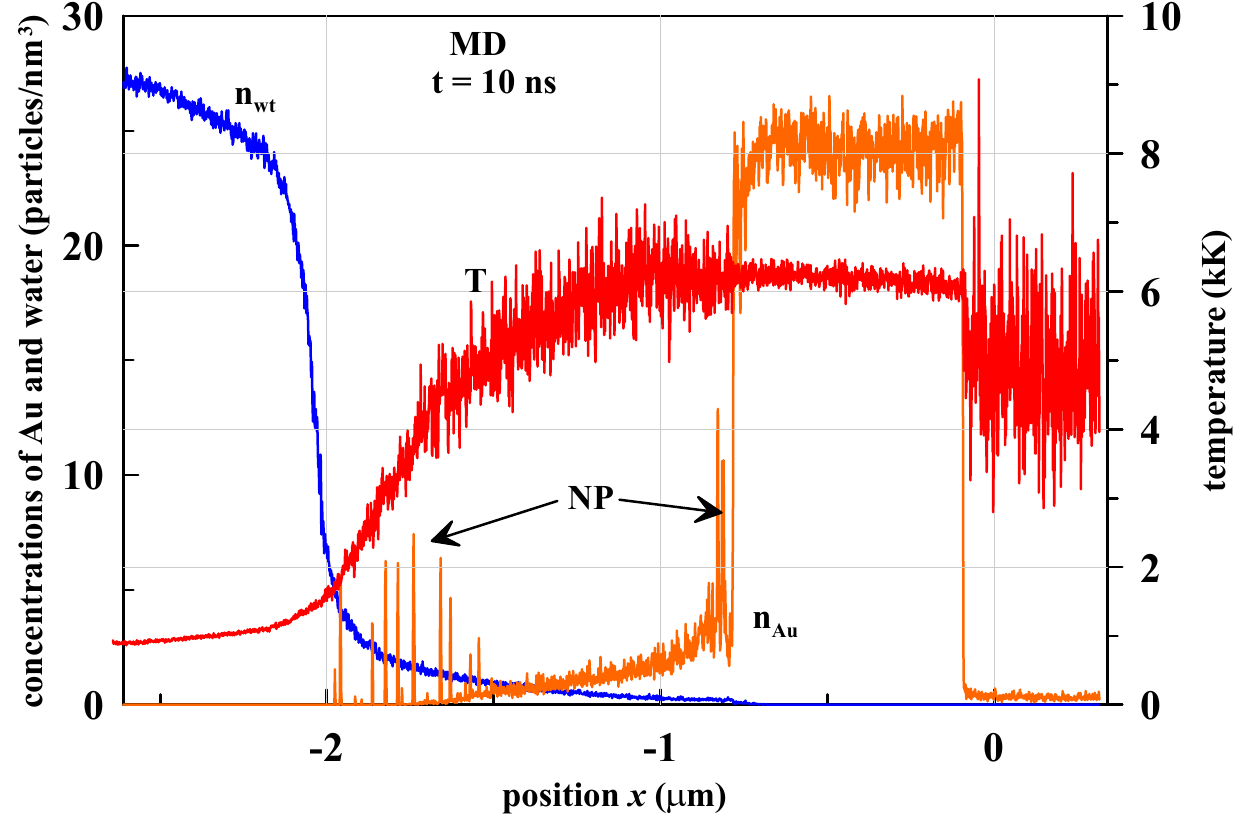}
\caption{\label{ris:08-concentrationsAu-wt+T} Previous Figure gives total density of water-gold mixture
 in the region of the mixture. Here we separate concentrations of gold $n_{Au}(x,t)$ and water $n_{wt}(x,t)$ for $t=10$ ns.
 Thus the diffusion layer and water-Au interpenetration becomes clear.
 The temperature profile "T" shows definite decrease into the water side.
 Thus it seems that diffusion overruns heat conduction to some extent.
 The two forests of the needles marked by the arrows "NP" (nanoparticles) present condensation process creating NPs.
    }  \end{figure}

 \section{Structure of high entropy layer according to molecular dynamics}

 Effective approach to solve problems related to laser ablation in liquids is based on molecular dynamics (MD) approach
  \cite{INA.jetp:2018.LAL,POVARNITSYN:2013,SHIH20173,LZ+Stephan:2018.LAL,INA.AIP:2018.LAL,INA.arxiv:2018.LAL,Ivanov2017,ANGELarXiv}.
 Hydrodynamic simulations presented above and MD compliment each other.
 Figures \ref{ris:07-HD-MD-densities} and \ref{ris:08-concentrationsAu-wt+T} illustrate the MD results.
 In Fig. \ref{ris:07-HD-MD-densities} the HD and MD density profiles are compared.
 The HD code doesn't include diffusion.
 Thus the HD contact is atomically sharp.
 From Fig. \ref{ris:08-concentrationsAu-wt+T} it is obvious that diffusion is important for nanoparticles formation.
 We have strong intermixing of gaseous gold and gaseous water in the case simulated by MD
    and shown in Figures \ref{ris:07-HD-MD-densities} and \ref{ris:08-concentrationsAu-wt+T}.
 There is the second drawback in the HD-code; the first is absence of diffusion.
 It relates to description of thermodynamics of gaseous gold.
 We use a sum of electron and ion free energies $F_e+F_i$ to obtain pressure.
 The particular expression for electron pressure $p_e>0$ has inaccuracy $\sim 0.01-0.1$ GPa.
 This inaccuracy is insignificant when pressures are at the GPA level.
 But due to this inaccuracy we overestimate total pressure $p=p_e+p_i$ of gaseous gold
    in the states with temperatures below $\approx 5$ kK.
 Additional simulations will be made in future with exclusion of this drawback.
 Then density of gaseous gold should be higher and thickness of the layer occupied by the gaseous gold should be more narrow.
 Comparison of the HD and MD simulations in Fig. \ref{ris:07-HD-MD-densities} says
   that nevertheless the real situation is approximately well described down to pressures $\sim 10^3$ bar.
 Estimates relating to the evolution at the much lower pressures are given in next Section.

 % 3-2

 Diffusion and heat conduction operate approximately at the same rate.
 This conclusion follows from Fig. \ref{ris:08-concentrationsAu-wt+T}.
 We see only narrow layer of pure hot water ahead the diffusion layer,
   this is the layer to the left from the digit "2" in Figure \ref{ris:07-HD-MD-densities},
     where density of water is decreased. (The layer of hot pure water is narrow relative to thickness of a diffusion layer.)
In Fig. \ref{ris:08-concentrationsAu-wt+T} this layer locates between
 the region of steep decrease of concentration of water molecules $n_{wt}(x,t)$
   and the first needle from the forest of needles at the left side of a diffusion layer.
 Let's mention that water at some distance from the diffusion layer (region 1 in Fig. \ref{ris:07-HD-MD-densities})
  is heated to temperatures near 1 kK, see Fig. \ref{ris:08-concentrationsAu-wt+T}.
 This heating is the dissipative trace of strong shock compression at the early stage in the case with an ultrashort pulse.

% 3-3

 Why diffusion is important for condensation?
 Pressure profile is approximately homogeneous along the layer of liquid water (region 1 in Fig. \ref{ris:07-HD-MD-densities})
   and in the diffusion layer; the last is filled with a mixture of gaseous water and gaseous gold.
Above we judge about the question is matter in the one-phase gaseous state or in the two-phase vapor-liquid state
  comparing pressure $p$ and vapor saturation pressure $p_{sat}(T)$ at given temperature:
   condensation begins if $p<p_{sat}(T).$ But this conclusion is valid for pure matter.
 In the case of mixture when we decide will gaseous gold condense or not we have to compare
   not total pressure of mixture $p_{total}$ with $p_{sat}$ but partial pressure of gaseous gold $p_{Au}$
    dissolved in water with $p_{sat};$ $p_{total}=p_{Au}+p_{wt}.$

% 3-4

 In the case shown in Fig. \ref{ris:08-concentrationsAu-wt+T} we have diluted solution of atomic gold
  at the left edge of the diffusion layer.
 Total pressure 1000 bar is rather high
  (it is above saturation pressure for this temperature - therefore condensation is delayed
    in the middle of the diffusion layer) in MD simulation at the instant $t=10$ ns
      shown in Figures \ref{ris:07-HD-MD-densities} and \ref{ris:08-concentrationsAu-wt+T}.
 But at the left edge of the diffusion layer partial pressure $p_{Au}$ is below $p_{sat}.$
 Also water particles act as a buffer gas which serve for withdrawal of latent heat of condensation
  in a pair Au-Au collisions thus enhancing rate of condensation.
 Therefore formation of nanoparticles begins and continues at the left edge.
 The condensed nanoparticles manifest themselves as the sharp needles at the density and gold concentration profiles
   in Figures \ref{ris:07-HD-MD-densities} and \ref{ris:08-concentrationsAu-wt+T}.
 They are marked by the left arrow in Fig. \ref{ris:08-concentrationsAu-wt+T} going from the note "NP" (nanoparticles).

% 3-5

 The second group of nanoparticles (group of the needles) appears and begins to grow at the right edge of diffusion layer.
 At this edge temperature is slightly lower while density of gaseous gold is few times higher than in the middle.
 Let's consider this situation (we compare the middle and the right edge) at the density-pressure phase diagram.
 Let's imagine the condensation curve $\rho_{sat}(p)$ at this plane $\rho,p.$
 Then the middle is outside the condensation curve $\rho_{middl}<\rho_{sat}(p)$
   while the right edge of the diffusion layer is inside $\rho_{right}>\rho_{sat}(p)$ because density is higher,
     here $p$ is pressure across the diffusion layer at the instant shown.
 Thus condensation and nanoparticles (needles) appear at the right edge - this place is marked by the right arrow going from the note "NP"
   in Fig. \ref{ris:08-concentrationsAu-wt+T}.

\begin{figure}       %  ---------------------------------------------  РИС. 9
   \centering   \includegraphics[width=0.75\columnwidth]{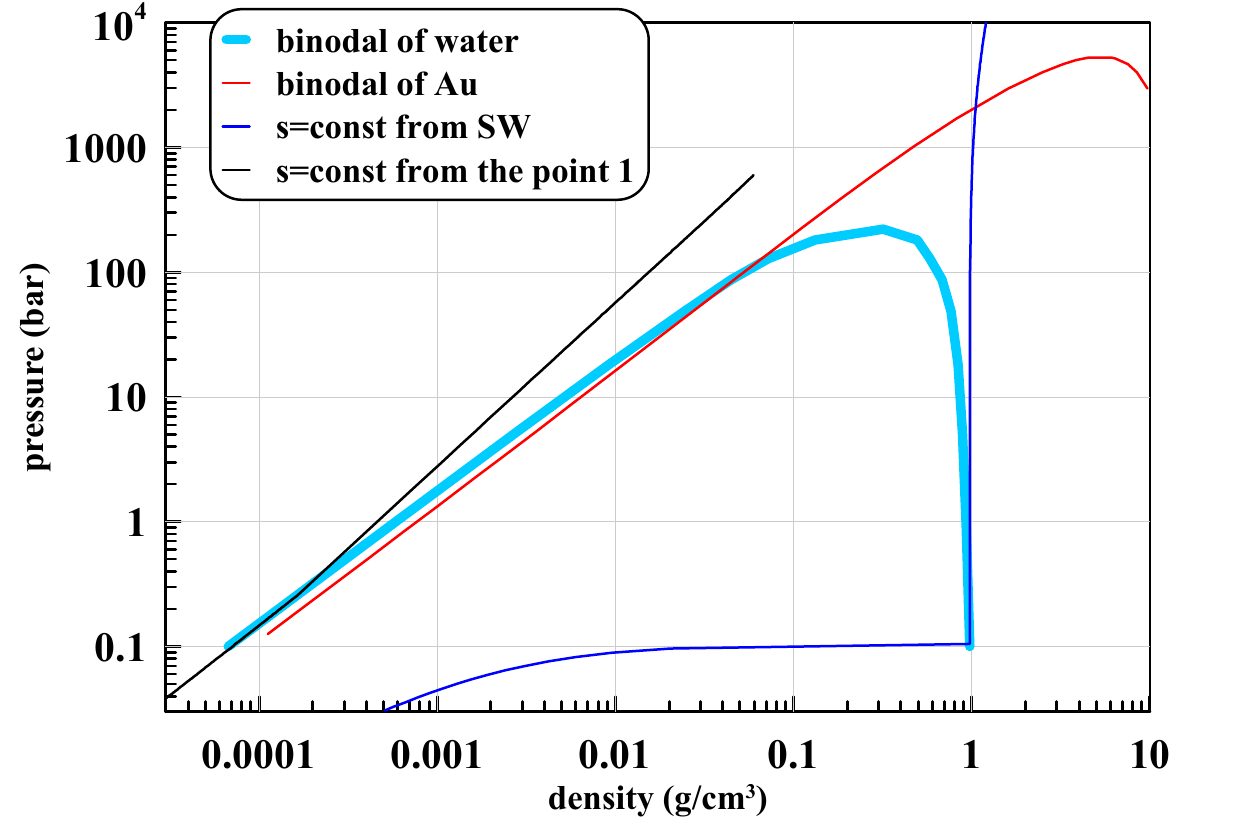}
\caption{\label{ris:09-s-curve} The density-pressure phase diagram is presented.
The thick light blue and red curves are the phase equilibrium curves for water and gold, resp.
Two adiabatic curves for water particles are shown.
One of them (the black curve) starts from the point 1 in Fig. \ref{ris:06-contactLayer}.
The other one (the deep blue curve) begins behind a shock front compressing water to 3 GPa.
    }  \end{figure}

\begin{figure}       %  ---------------------------------------------  РИС. 10
   \centering   \includegraphics[width=0.75\columnwidth]{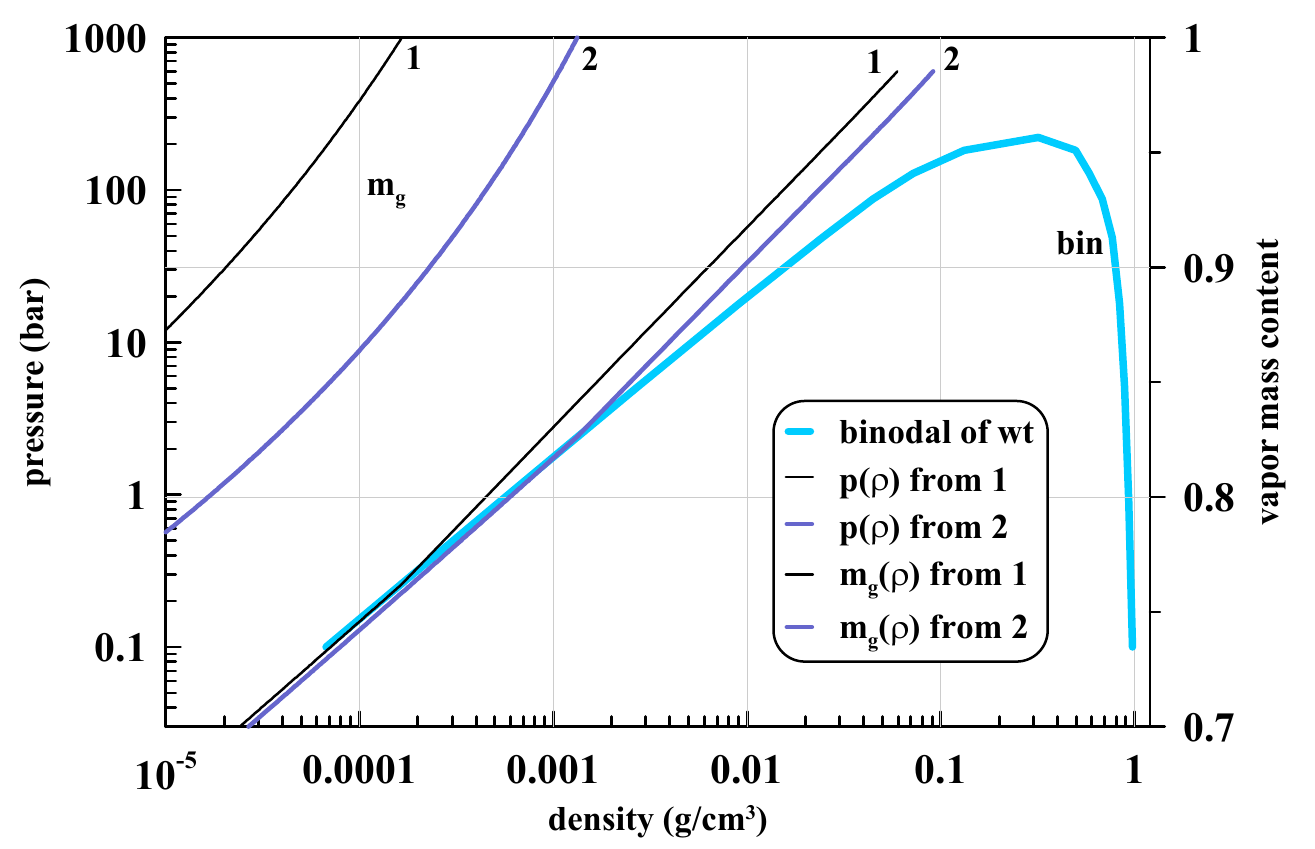}
\caption{\label{ris:10} The coexistence curve (bin) and adiabatic curves 1 and 2 are shown.
 The starting points of these curves are the points 1 and 2 in Fig. \ref{ris:06-contactLayer}.
 The pair of the curves 1 and 2 belongs to the plane $\rho,p$ while the another pair 1 and 2 belongs to the plane $\rho,m_g.$
 The $m_g$ is the mass fraction of gas in the vapor-liquid mixture.
 It is presented at the right vertical axis.
    }  \end{figure}

% 3-6

 Hydrodynamic (HD) code used is described in our previous papers
  \cite{INA.jetp:2018.LAL,INA.AIP:2018.LAL,INA.arxiv:2018.LAL,ANGELarXiv}.%,ICPEPAarXiv}.
 In this paper as was said above we add conductivity of water to the HD code.
 Density profile obtained from the HD code is given in Figures \ref{ris:05-contactLayer} - \ref{ris:07-HD-MD-densities}.
 The HD simulated case relates to the long (nanosecond) pulse.
 It is important that in this case we don't see nucleation, formation of foam, breaking of foam
  and separation of the right side of the gaseous-foamy layer from the bottom of the future crater;
   we say here "the right side" because laser beam comes from the left side as shown in our Figures.

 % 3-7

 The MD simulation relates to the ultrashort laser pulse. In this case we have foaming, breaking and separation of foam.
 Descriptions of these stages needs separate text.
 Here in Figures \ref{ris:07-HD-MD-densities} and \ref{ris:08-concentrationsAu-wt+T} we show the rather late stage
   achieved during our simulation. Formulation of problem solved by the MD code is following.
 We have a long computational cell with 2 microns of water and 2 microns of gold
  and $10\times 10$ nm$\times$nm cross section,
   see also descriptions in \cite{INA.jetp:2018.LAL,INA.AIP:2018.LAL,INA.arxiv:2018.LAL,ANGELarXiv}.%,ICPEPAarXiv}.
 We did preliminary simulations defining trajectory of water particle $x_{400}(t)$
  located before a laser pulse rather outside (400 nm outside) to the initial position of the water-gold contact.
 The trajectory is defined along a range of times up to the instant when a rarefaction wave
   reflected from the free surface of water (free surface is placed initially at the distance $-2$ microns
    from initial contact) comes to our marked water particle $x_{400}.$
 After that we use this trajectory and its analytical continuation in time as the left boundary of the computational cell
   to decrease number of atoms in simulation.

 % 3-8

 We follow evolution of a thick gold target up to the stage when the foam separates from the bottom of a crater.
 This takes place at the depth $\approx 450$ nm below (to the right side) from initial position of a contact
   - the point $x=0$ in our Figures - for the MD shot presented here.
 After that we delete atoms of gold to the right from this bottom because, indeed,
   the approximately motionless condensed part of a rest of a gold target play insignificant role
     in evolution of a separated plume.
 We put the right boundary condition at this depth. The condition deletes gold atoms which achieve this depth.
 Then gradually formation of the clot 4 and the rarefied region 5
  in Figures \ref{ris:07-HD-MD-densities} and \ref{ris:08-concentrationsAu-wt+T} takes place.

 % 3-9  замороз поверхнстн структуры происходит ДО окончания эвол плюма

 Evaporated/ablated mass of gold in HD simulation of a nanosecond action is $\approx 20$ nm
   in initial density 19.3 g/cm$\!^3$
 While in the case of the ultrashort action with larger energy $F_{abs}$ this mass is $\approx 400-500$ nm;
   see Figures \ref{ris:07-HD-MD-densities} and \ref{ris:08-concentrationsAu-wt+T} where the point $x=0$
     corresponds to the initial position of a contact in the HD and MD cases both.
 Cooling and recrystallization in the condensed residuals of the gold targets in the both cases proceeds faster
   than evolution of the plume.
 Therefore the plume evolutions are very different in the cases with ablation in vacuum or gas versus ablation in liquid
   - this was emphasized in several papers
 \cite{POVARNITSYN:2013,SHIH20173,LZ+Stephan:2018.LAL,INA.jetp:2018.LAL,INA.AIP:2018.LAL,INA.arxiv:2018.LAL,ANGELarXiv}.%,ICPEPAarXiv}.
 While the solidification process and thus formation of surface structures are much less dependent on the environment
  (vacuum or liquid), e.g., the random surface structures produced by ultrashort pulse are the same
    in the cases with ablation in vacuum and in liquid.
 This conclusion follows from analysis of dynamics of the plumes given above.

\begin{figure}       %  ---------------------------------------------  РИС. 11
   \centering   \includegraphics[width=0.75\columnwidth]{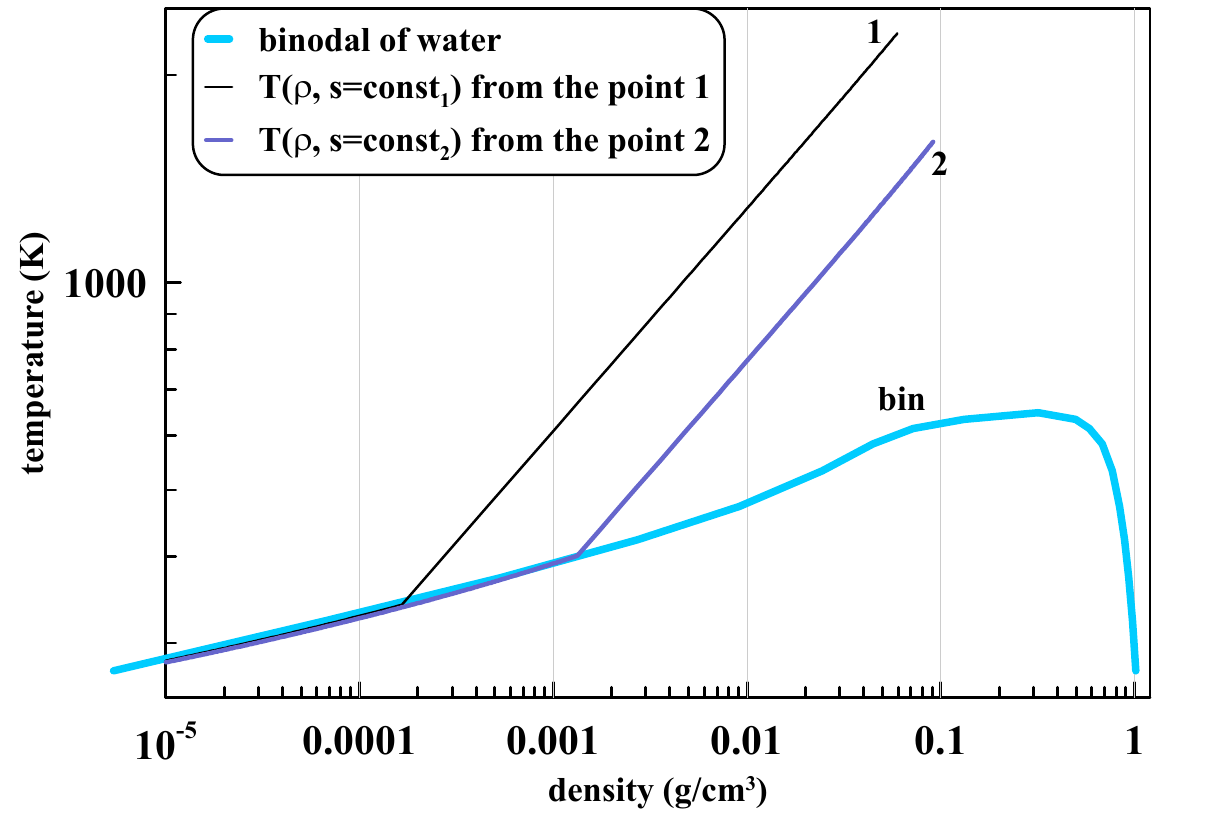}
\caption{\label{ris:11} The same as in Fig. \ref{ris:10} but now at the $\rho, T$ phase plane.
    }  \end{figure}

 % 3-10     % сструктр укли   ссылки    что первые мы объяснили физику происхождния случнх структр в духе Воробьева
 % структ нс-импульс  ссылки=??

 Appearance of the random surface structures (RSS) after ultrafast single shot action
   was demonstrated in papers by Vorobyev and Guo \cite{Vorobyev:06,Vorobyev:2008}.
 First explanations connecting the RSS and foaming/rupture of foam
  and fast (before their disappearance thanks to surface tension) freezing of the cellular capillary remnants of foam
   (the remnants are: membranes, jets and droplets) were given in the papers
   \cite{Zhakhovskii2008a,Zhakhovskii2008b}
    later confirmed in the papers
     \cite{Ashitkov:QuantEl:foam:2014, Inogamov:2014:subSurfVoids:JPCS, Inogamov:2014:subSurfVoids:JPCS, icosahedral:2016,
     Ionin:2016, Zhigilei:SurfuceNanoStrc:2018}.
 These results relate to the RSS observed after single shot ablation in vacuum.
 According to conclusion given in the previous paragraph the same RSS structures
  should be formed in the cases with ablation in liquid initiated by an ultrashort pulse.
 Indeed, previous observations (see Fig. 8 in \cite{Shafeev:2009Greece} and \cite{Barmina:SurfStructr:2009})
   and recent direct comparisons of RSS in gas and in liquid
    (see Fig. 2 in \cite{Kudryashov:SurfStruktr:air:liquid:2018} and \cite{SARAEVA20191018})
     confirm the conclusion.
 Weakly pronounced structures form in the case of the nanosecond pulses.

 \section{Late stages, low pressures}

 % 4-1

 At the present stage of studies, we haven't longer simulations than shown above.
 Future work will be continued using transport coefficients for hot rarefied matter like those given in the papers \cite{ClerouinRakhel:2008,FrenchRedmer:2010}.
 It is plausible that at the late stages the conductive cooling is slow.
 Then it is worth to follow an adiabatic expansion up to very low pressures.
 Let's take the points 1 and 2 from Fig. \ref{ris:06-contactLayer}. These points correspond to pure water.
 Their coordinates at the instant shown in Fig. \ref{ris:06-contactLayer} are
       $\rho=0.06$ g/cm$\!^3,$ $T=2.3$ kK for the point 1
   and $\rho=0.09$ g/cm$\!^3,$ $T=1.6$ kK for the point 2, and both points are under pressure $p=600$ bar
     because a pressure field is approximately homogeneous across the layers of hot water and gaseous gold.
 The adiabatic curve (s=const from the point 1) starting from the point 1 is shown in Fig. \ref{ris:09-s-curve};
   it is the black curve.
 Both adiabatic curves (black and deep blue) are shown in Figures \ref{ris:10} and \ref{ris:11}.

 % 4-2

 We follow inflation of water along the adiabatic (black) curve in Fig. \ref{ris:09-s-curve}
  and along the curves 1 and 2 in Figures \ref{ris:10} and \ref{ris:11} down to expansion degree
   below the ambient pressure in water which is chosen equal to 1 bar.
 The curve "s=const from the point 1" in Fig. \ref{ris:09-s-curve} consists from the two parts
   corresponding to one- and two-phase states.
 Their indices $\gamma=d(\log(p))/d(\log(\rho))$ are 1.3 and 1.06, resp.
 Compare these values with the values 1.3-1.16 given in Fig. 4 in \cite{Amans:2016:APL}.
 Only at very small densities the black curve intersects the binodal curve for water.

 % 421

 The curve 2 in Figures \ref{ris:10} and \ref{ris:11} also consists from the one- and two-phase parts.
 The two-phase part of the adiabatic curve begins when the curve intersects the coexisting curve named binodal
   - this is the light blue, thick curve in Figures \ref{ris:09-s-curve}-\ref{ris:11}. It relates to water.
 The curves 1 and 2 have the kink points where their slope is changed to finite value.
 The kink points are the intersection points.
 The kink is especially definitely seen at the phase plane $\rho,T$ shown in Fig. \ref{ris:11}.

 % 421.2

 Liquid water content appears in dry water gas after the intersection point
   when the adiabatic curve transits into a two-phase region.
 Two-phase mixture consists from vapor and liquid.
 The mass content of vapor as function of average density of a mixture is shown in Fig. \ref{ris:10} -
    the curves $m_g$ number 1 and 2.
 We see that there are rather large mass of liquid water in the mixtures.
 At the same average density of a mixture the water content is larger if entropy is smaller.
 In the limit $T\to 0$ a mixture totally condensed into liquid.
 In the cases considered, the 2-phase adiabatic curves locates very near to the coexistence curve.

 % 4-3

 The adiabatic curves for gaseous water filling a semi-bubble are concentrated near the hottest curve
   having maximum entropy.
 In the case of a long pulse (when shock and its dissipative heating are weak) they belong
   to the conductively heated layer.
 The adiabatic curve starting from a water particle heated by 3 GPa shock has small degree of expansion
   at final pressure 0.1 bar.
 This is the deep blue curve "s=const from SW" in Fig. \ref{ris:09-s-curve}; SW is shock wave.

 Writing the mass balance between a disk of hot water and a semi-bubble
  \begin{equation} \label{eq:03-massBalance}
 \pi R_L^2 d_{wt} \, \rho_1 = (1/2) \, (4/3) \, \pi R_{bub}^3 \, \rho_{fin},
  \end{equation}
  we find that the maximum radius (when the minimum $p$ is achieved) is $R_{bub}\approx 800$ $\mu$m.
 This value is less than the $R_{bub}\approx 1.4$ mm measured in \cite{Amans:2016:APL}.
 In (\ref{eq:03-massBalance}) we assume that $R_L=250$ $\mu$m as in \cite{Amans:2016:APL},
  $\rho_1=0.06$ g/cm$\!^3$ is density in the point 1 in Fig. \ref{ris:06-contactLayer},
   $\rho_{fin}=6\cdot 10^{-5}$ g/cm$\!^3$ is average density of two-phase water at the end point of the black curve in Fig. \ref{ris:09-s-curve}.
 We put thickness of a layer of hot water equal to $d_{wt}=5$ $\mu$m according to (\ref{eq:02-dWaterFinal}).

 \section{Conclusion}

 We have considered laser ablation of gold in water by pulses of different durations and absorbed energies.
 Estimates based on Stokes's law show that we can neglect creep of gold nanoparticles (NP) with diameters less than ten nm
   relative to gaseous water surrounding them.
 We show that NPs are mixed with gaseous water in the layer of hot water, see Fig. \ref{ris:08-concentrationsAu-wt+T}.
 Thus the NPs fill a semi-bubble of gaseous water.
 Their concentration and sizes at different radii differ.
 The outer and inner NPs are created earlier relative to the NPs filling the middle range of radii inside a semi-bubble,
   see Fig. \ref{ris:08-concentrationsAu-wt+T}.
 Successive expansions and collapses of a semi-bubble mix this spatially differentiated distributions.

\bibliographystyle{elsarticle-num}
%\bibliography{GoldLAL}
%\bibliography{Au-glass_50ps_Vitya_28.11.18.bib}

%
 \end{document}